\documentclass[iop]{emulateapj}
\usepackage{aabib}
\def\rd{Di\thinspace Stefano}

\def\actaa{AcA}

\slugcomment{Accepted 2012 December 27 for publication in Astrophysical Journal}

\shorttitle{Nearby Planetary Systems As Lenses}
\shortauthors{Di Stefano, Matthews, L\'epine } 

\begin{document}
\bibliographystyle{aabib}
\title{Nearby Planetary Systems As Lenses During Predicted 
Close Passages to Background Stars}
\author{Rosanne Di\thinspace~Stefano\altaffilmark{1}, James
  Matthews\altaffilmark{1,2}, and S\'ebastien L\'epine\altaffilmark{3,4}}
\altaffiltext{1}{Harvard-Smithsonian Center for Astrophysics, 60 Garden
  Street, Cambridge, MA 02138}
\altaffiltext{2}{School of Physics \& Astronomy, University of Southampton,
  Southampton, SO17 1BJ, UK}
\altaffiltext{3}{Department of Astrophysics, Division of Physical Sciences,
  American Museum of Natural History, Central Park West 79th Street,
  New York, NY 10024}
\altaffiltext{4}{Department of Physics, Graduate Center, City University of New York, 365 Fifth Avenue, New York, NY 10016, USA}

%%%%%%%%%%%%%%%%%%%%%%%%%%%%%%%%%%%%%%%%%%%%%%%%%%%%%%%%%%%%%%%%%%%%%%%%%%%
%
%                   ABSTRACT
%
%%%%%%%%%%%%%%%%%%%%%%%%%%%%%%%%%%%%%%%%%%%%%%%%%%%%%%%%%%%%%%%%%%%%%%%%%%%

\begin{abstract}
The Einstein rings and proper motions of nearby stars 
tend to be large. Thus, every year some foreground stars within
a few hundred parsecs of Earth  
induce gravitational lensing events in background stars.
In some of these cases, the events may exhibit evidence of planets
%=> the events may exhibit
orbiting the nearby star. In fact, planets can even be discovered 
during relatively distant passages.
Here, we
study the lensing signatures associated with planets
orbiting nearby high-proper-motion stars.
We find the following. (1)~Wide-orbit planets 
can be detected for all distances of closest approach between the 
foreground and background stars, potentially producing independent events   
long before and/or after the
closest approach. (2)~Close-orbit planets
can be detected for intermediate distances of closest approach,  
producing quasiperiodic signatures that may occur 
days or weeks
before and after the stellar-lens event. (3)~Planets in the so-called
``zone for resonant lensing'' can significantly increase the
magnification when the distance of closest approach is small, 
making the stellar-lens event easier to detect, while
simultaneously providing evidence for planets. 
Because approaches close enough to allow planets to be detected can be predicted,
we can plan observing strategies to take advantage of the theoretical framework
built in this paper, which describes the sequence of expected effects
in terms of a sequence of detection regimes. 
\end{abstract}

\maketitle

\def\rd{Di\thinspace Stefano} 

\section{Introduction} 

We are entering an era in which we can systematically predict close passages
between nearby stars moving across the sky and background stars.
Predictions are now possible because 
%more than $2$~million nearby stars 
%have well-measured proper motions (L\'epine, private communication),
 more than 2 million nearby stars with the largest motions have
 been identified (L\'epine \& Shara 2005, L\'epine \& Gaidos 2011),
while deep surveys are mapping large
background fields 
%(Kaiser 2004; {\bf more refs}).
 (Kaiser 2004; Aihara et al. 2011)

During the closest passages, the combined gravitational influence
of the nearby star and its planets can produce distinctive lensing events
which reveal the presence of the planets.
In fact, even when the nearby star
itself does not come close enough to the background star to produce detectable
lensing effects, it should still be possible to probe for the presence of planets.  
In this paper we develop the theoretical framework needed to
take full advantage of the planet-lens opportunities afforded by close passages 
between nearby and more distant stars. 

For individual cases in which a close
passages is predicted, 
we consider the
possibility that the nearby star making the close passage has planets in 
orbits with semimajor axes ranging outward from about a 
tenth of an Einstein radius.
For each orbital size, we predict the full range of detectable signatures
as a function of time to the point of closest approach. 
%We consider the general case and also derive insight from considering a
%specific example, the close passage between VB~10 and a background star
%that occurred during late 2011/early 2012. 
 In a future paper, based on \rd , Matthews, \& L\'epine (2012), 
we turn to the
design of observing programs that are optimized to detect the
signatures predicted we predict here.

Section 2 provides a brief overview of the
relevant lensing background. 
In \S 3 we also discuss other relevant characteristics of planetary systems.
Of particular importance is the orbital period. Because stars making predicted 
passages  will generally be nearby, their Einstein radii can be small.
Thus, many of the planets that can be discovered through lensing may
%=> planets that can be discovered
have short orbital periods. The change in orbital phase during the interval
of  close passage can increase the probability of planet detection.
In addition, the reflexive orbital motion of the central star
can effectively be magnified, providing another avenue toward the 
detection of planets. We also consider the possibility that the planets to 
be detected will lie in the habitable zone. 

The types of lensing signatures associated with planets
depend strongly on the orbital separation. It is convenient to
divide the full range of separations into three subranges or zones: 
the close-orbit zone,
the ``resonant'' zone, and the wide-orbit zone. 
Planets in each zone produce distinctive signatures. The signatures
produced by planets in a given zone can, however, be detected 
%=> zone can, however, be
only if the background source passes within
a well-defined range of distances of closest approach. 
We therefore define, for each
zone, a detection ``regime'' of distances of closest approach: when the source
enters the regime associated with a particular ``zone''  
of planetary positions, we have a significant chance of detecting any
planets that may inhabit that particular zone around the nearby star.

By utilizing the notion of three separate lensing regimes, we can
organize the effects according to their likely times of occurrence.
We begin this process in Section 4, where we consider the regimes in the 
order of increasing proximity to the central star, which is also
the order in which planets can be detected: wide, close, and then resonant.  
In Section 5 we summarize our results and 
prepare the way for their use in real observing programs, which will be
the subject of a future paper, based on \rd , Matthews, \& L\'epine (2012).  

\section{Background: From Close Passages to Lensing Events}

Given enough high-resolution images of a region, the relative motion of the
stars it contains can be measured and close passages can be predicted.
We are interested in the question of whether
close passages will produce lensing events (Di Stefano 2008a, 2008b).
The simplest case to consider is lensing of 
 a distant background star by a star much closer
in the foreground.

Gravitational lensing has both photometric and astrometric effects
(e.g., Einstein 1936; Dominik \& Sahu 2000).
One clear-cut astrometric effect is illustrated by the concept of the
Einstein angle, $\theta_E$.
The Einstein angle is defined to be 
the angular radius of the ring that would form the
image of the source, were the source, lens, and observer to be
perfectly aligned. 
In the absence of perfect alignment, a point lens
produces two images. 
If $M_\ast$ is the mass of the lens, $D_L$  is the 
distance to the lens, and $D_S$ is the distance to the lensed source, then  
\begin{equation}  
\theta_E = 10\, {\rm mas}\,  
 \Bigg[\Big(\frac{M_\ast}{0.1\, M_\odot}\Big)\, 
                         \Big(\frac{8\, {\rm pc}}{D_L}\Big)\, 
                         \Big(1-\frac{D_L}{D_S}\Big)\Bigg]^\frac{1}{2}. 
\end{equation}
%=> [Would it not be simpler to just replace ``x'' by DL/DS in this 
%=> and all other equations, and just not use ``x'' at all in
%=> the manuscript?
By comparing the size of the Einstein angle to the distance of
closest approach, we can determine how large the effects of lensing will be.
It is therefore important to be able able to estimate the value of 
$\theta_E.$ 

Prediction of close passages is most likely to be possible when the
nearby star is a high-proper-motion star (HPMS), with 
proper motion $\mu$ larger than a few
tens of milliarcseconds per year. Such stars are near enough to
us that    
the value of $\sqrt{1-D_L/D_S}$ is likely to be close to unity. 
We can therefore
often obtain a useful estimate of the value of $\theta_E$ if we can 
estimate the mass of the nearby star and its distance from us.  

Because we are considering nearby stars with measured proper motions, in most cases we will know the star's spectral type and can use it to 
estimate the values of $D_L$ and $M_\ast$. For some nearby star we can do
even better, because the geometric parallax is known and provides a
high-precision measurement of $D_L.$ 
VB~10 is one such well-studied star, with an estimated distance of 
$5.8$~pc and an estimated mass of $0.075\, M_\odot.$ The Einstein angle for
VB~10~ is approximately equal 
 to $10$~milliarcseconds (L\'epine \& DiStefano 2012).

The astrometric effects of lensing fall off slowly  
with 
the angular separation between the source and lens, $u$.
If $u$ is 
expressed in units
of $\theta_E,$ then in the limit of large $u,$  
the shift  of the centroid is roughly equal to $\theta_E/u.$ 
 Prospective space-based missions, such as GAIA and Sim-Lite
 potentially could measure
centroid shifts
of tens to hundreds of microarcseconds, and may therefore be able to
detect the effects of lensing event when the distance of closest 
approach is as large as an arcsecond or more.
Measuring astrometric shifts provides the advantage of a direct measurement
of $\theta_E.$ If, therefore, $D_L$ is well measured, the gravitational
mass can be measured to high precision. 
 Such a direct, high-precision measurement can have
 important scientific relevance.
For example, the estimated mass of VB~10 
lies close to the boundary between stars and brown dwarfs, so that
a mass measurement 
would provide a particularly valuable data point. Other 
nearby objects for which direct mass measurements are potentially important 
include brown dwarfs and 
white dwarfs.  

In this paper we will concentrate on the photometric effect of lensing, i.e.,
the magnification of the background source. 
A unique value of the magnification is associated with 
each value of $u;$ $A=A(u)$. 
The fractional magnification,
$A-1,$
is roughly $0.34$ for $u=1,$ $0.06$ for $u=2,$ and $0.01$ for $u=3.5.$
For large $u,$ the magnification falls off as $1/u^4$.
Thus, if, for example, an event with a peak magnification of $2\%$  can be
reliably identified, then the required angle of closest approach is
$\sim 3\times \theta_E;$ if passages this close can be
predicted, then we can be assured of detecting lensing of the background
star by the foreground star.

A lensing model fit to the magnification light curve 
allows us to derive the value of the Einstein diameter crossing time, 
$\tau_E = 2\, \theta_E/\mu,$ 
where $\mu$ is the proper motion. The value of $\mu$ can be measured
for nearby stars, as it has been for VB~10.  
With $\tau_E$ measured from the light curve, the value of $\theta_E$
can be determined from the event itself, without the need for astrometric
measurements.  

\section{The Effects of Planets}

Because we are interested in discovering planets orbiting the nearby star,
the orbital separation between the planet and its star, generally
expressed in units of the star's Einstein radius, will play a 
key role. It is therefore important to introduce the Einstein radius,
$R_E = D_L \theta_E.$
\begin{equation}  
R_E = 0.08\, {\rm AU}\,  
 \Bigg[\Big(\frac{M_\ast}{0.1\, M_\odot}\Big)\, 
                         \Big(\frac{D_L}{8\, {\rm pc}}\Big)\, 
                         \Big(1-\frac{D_L}{D_S}\Big)\Bigg]^\frac{1}{2}, 
\end{equation}
%=> [Replace ``x'' by DL/DS?]
If $a$ is the instantaneous projected separation 
between a nearby star and a specific
planet it may harbor, then $\alpha=a/R_E$ is the quantity that determines
the types of light curve features that can provide evidence that the
planet exists.  It is therefore useful to  
delineate certain ``zones'' around the stellar lens, 
each zone defined by a range
of values of $\alpha$.

\subsection{Orbital Periods}
Equation (2) demonstrates that, for planets that are close to us, 
the Einstein radius,  
$R_E=D_L\, \theta_E$,  
can be relatively small\footnote{The Einstein radius scales as 
$D_L^\frac{1}{2}$ for nearby lenses, while the size of
the Einstein ring, $\theta_E,$ scales as 
$D_L^{-\frac{1}{2}}$ for nearby lenses}. 
The exact value of $R_E$ depends on both the lens mass
and on the value of $D_L.$ Nevertheless, for a wide range of stellar masses and
distances to the lens (within a few hundred pc), $R_E$ can be smaller
than an AU. 

For a face-on circular orbit, the value of $\alpha$ is constant and is equal
to the semimajor axis. In this case, 
\begin{equation} 
P_{orb}=(26.1~{\rm days})\, \alpha^\frac{3}{2} 
\Bigg(\frac{M_\ast}{0.1\, M_\odot}\Bigg)^\frac{1}{4}  
\Bigg(\frac{D_L}{8\, {\rm pc}}\Bigg)^\frac{3}{4}  
\Bigg(1-\frac{D_L}{D_S}\Bigg)^\frac{1}{4} 
\end{equation}
In general, the orbit will not be face-on, and may also be
elliptical. Thus, $\alpha=\alpha(t)$, and the semimajor
axis is $p(t)\, \alpha(t)$, where   
the instantaneous value of $p(t)$ can be 
larger than or smaller than unity,
depending on the orbital inclination, eccentricity, and phase.
 For the purpose of considering specific well-defined
cases, we will use circular face-on orbits as examples.
We note however, that when deriving models to fit 
observed light curves, the full range
of possible orbits must be considered.

Short orbital periods mean that the orbital phase can 
change significantly during
the interval when the lens is close to the source. 
For example, if $R_E = 0.08$~AU,
then the time taken by a $10$~km~s$^{-1}$ lens
to move through a distance of $2\, R_E$ is $27.7$~days. Thus, for a  
lens with $M_\ast=0.1\, M_\odot,$ located $8$~pc away, 
a full orbital period can
occur (if the orbit is circular and face-on, with $\alpha=1$)
during the time required by the lens to move across an Einstein diameter.
More revolutions can occur if $\alpha$ is smaller. In fact, because the ratio
of $P_{orb}$ to $R_E$ is proportional to $D_L^\frac{1}{4},$ and  
to $M_\ast^{-\frac{1}{4}},$  
there are wide ranges of parameters for which orbital motion is important.

It is therefore
more likely that regions in which the isomagnification contours are perturbed
from the point-lens form will pass in front of the source, increasing
the probability of planet detection. 
In some cases, the region in the lens
plane with magnification deviations can pass
over the source star more than once. The repetition of deviations
makes them more likely to be detected and correctly identified. 

Overall, the probability of planet detection can be significantly
 larger for HPMSs
than it is for
lens stars located several kpc away, the typical location for the lenses
discovered by the lensing teams [e.g., the OGLE (Udalski 2003) and MOA
(Bond et al. 2001) teams, both of which are presently active].
%=> [Add references here to those lensing programs?]

\subsection{Magnification of Reflexive Motion}

As a planet orbits a star, the star wobbles in its own
smaller orbit. When the star serves as a lens, small changes in its position 
can make a measurable difference in the magnification it produces in
a background star. The magnification enhances the effects of stellar wobble,
making it more readily detectable.  

In units of the Einstein angle, the wobble of the star has amplitude
\begin{equation}   
\delta= \frac{m_p}{M_J} \, {\alpha}\times f,  
\end{equation}  
where $M_J$ is the mass of Jupiter, and the value of $f$ depends on how much of the orbit is executed during the
time when the magnification is detectable; 
if a full orbit is executed $f$ could be as large as about $2$,
or much smaller, if the shift in orbital phase during the event is small.

Whether the stellar wobble is large enough to produce a detectable change in
magnification depends on how close on the sky the source star is to the lens.
Let $\beta$ represent the projected distance between source and lens,
expressed in units of the Einstein radius.  
The magnitude variation can be expressed as $(A(\beta-\delta)-A(\beta+\delta))$.
The value of $\delta$ will generally be smaller than $\theta_E$. For close
approaches, where the reflexive motion is most magnified, the value of $\beta$ is also
small. Thus, the magnitude of the fractional change,
$\Delta A/A$ in magnification
can be written as follows.
\begin{equation}   
\frac{\Delta A}{A}= \Bigg|\frac{\beta\, \delta}{\beta^2-\delta^2}\Bigg|   
\end{equation}
For $\delta$ smaller than $\beta$, we may have $\Delta A/A=\delta/\beta.$ 
Thus, $\Delta A/A$ may well correspond to a measurable change in the magnification.
For close orbits (small values of $\alpha$), this change will be small,
but it will be periodic. For large $\alpha,$ the change may be large, but it
will not repeat. Instead, it can be measured by a shift in the peak magnification
relative to the baseline.

 \subsection{The Habitable Zone} 

The planets many astronomers are most eager to discover are those that
could harbor life. It is also desirable that lensing 
discovers planets that could be habitable 
(Di\thinspace Stefano 1999).
\rd\ \& Night (2008) pointed out that, when the lens is a nearby
dwarf star, planets that can produce detectable lensing signatures are more
likely to lie in the zone of habitability. This should apply to
the HPMSs for which we can predict events. 

The zone for habitability is a concept that is based on the
premise that a planet orbiting a star is more likely to harbor life
when its surface temperature is such that it can support
liquid water. This will be the case, when the flux incident on the
planet from the star is comparable to the flux received by the Earth from the 
Sun. This condition on the flux is satisfied if   
\begin{equation} 
D_L \, \Big(1-\frac{D_L}{D_S}\Big) = \frac{125\, {\rm pc}}{\alpha^2}\, 
\Bigg(\frac{M}{M_\odot}\Bigg)^{2.5}  
\end{equation}

The habitable zone is an annulus that is large enough that some systems
may have values of
$D_L$ and $M_\ast$ roughly compatible 
 with the habitable zone for a broad range of $\alpha$ values.
 In this case,
one close passage might be able to 
discover multiple planets that are potentially habitable.  

\section{Zones and Detection Regimes}

\begin{figure*}
\label{caustic}
\centering
\includegraphics[width=0.68\textwidth]
{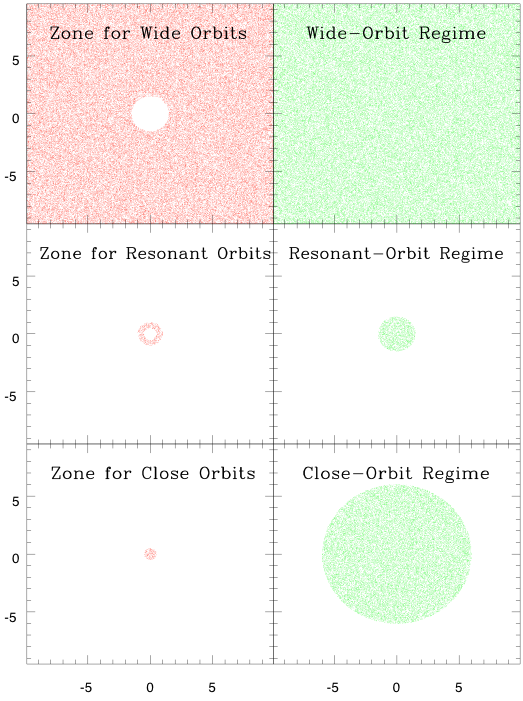}
%\vspace{-.5 true in} 
%created in /home/rd/meso 
\caption{{\footnotesize 
Panels on the left show the zones within which planets can be located. 
{\sl Bottom panel:} planets in the close-orbit zone have $\alpha<0.5$. 
{\sl Middle panel:} planets in the zone for resonant lensing 
have $0.5 <\alpha<2.$ {\sl Top panel:} planets in the wide-orbit 
zone have $\alpha>2$. Panels on the right show the corresponding 
detection regime. In each case, the detection regime contains the zone where
the planets are located, but also a larger region. {\sl Bottom panel:} The
close-orbit detection regime extends to just over 
$\frac{1}{\alpha_{min}}-\alpha_{min}$,
where $\alpha_{min}$ is the minimum value of $\alpha$ for which the
planet-induced deviations are detectable. 
Here we have taken $\alpha_{min}\sim 0.15$. {\sl Middle Panel:} The 
detection regime for resonant-orbit planet includes the resonant zone itself and
is augmented by a disk around the stellar position. {\sl Top Panel:} The
wide-orbit detection regime extends from the origin to the widest
planet in the planetary system, which can be at a distance of
hundreds of Einstein radii.}} 
\end{figure*}
 
The characteristic signatures 
associated with planets are different for different
orbital separations between the planet and the star. It is therefore 
convenient to define three distinct {\sl zones}  
in which planets may be located: the close-orbit zone, the
resonant-orbit zone, and the wide-orbit zone.
Each zone is defined by a range of values of $\alpha=a/R_E,$
representing the instantaneous value of the projected orbital separation
between the planet and its star, expressed in units of the 
Einstein radius.

In order to detect planets located in a given zone, the source 
must pass through a region in which there are 
planet-produced perturbations
in the magnification pattern. 
For planets in each zone, there is an associated
region where the deviations are most noticeable. The source track must pass 
behind this perturbed region in order for the presence of the planet
to be detected.  
We call this region the {\sl detection regime} 
associated with the planet-containing zone.
The detection regime is defined by the range of distances of closest
approach compatible with event detection. As in \S 3, it is convenient to
let $\beta$ represent the distance of closest approach,
expressed in units of $R_E.$  

As we will see,  the detection regime for each zone
contains the zone, but is larger. 
 The zones and their associated regimes are described below and are
illustrated in Figure~1.
We note that the boundaries between the zones containing 
each category of planetary orbits
are fuzzy, in the sense that there are generally
 not abrupt change in the signatures
at specific values of the projected orbital separation, and the signatures
also depend on the mass ratio. 
Nevertheless, the signatures change in a significant and systematic
way as $\alpha$ increases, making it useful to delineate the three zones
we discuss below, and the associated detection regime associated
with each.

\begin{figure*}
\label{caustic}
\centering
\includegraphics[width=1.0\textwidth]
{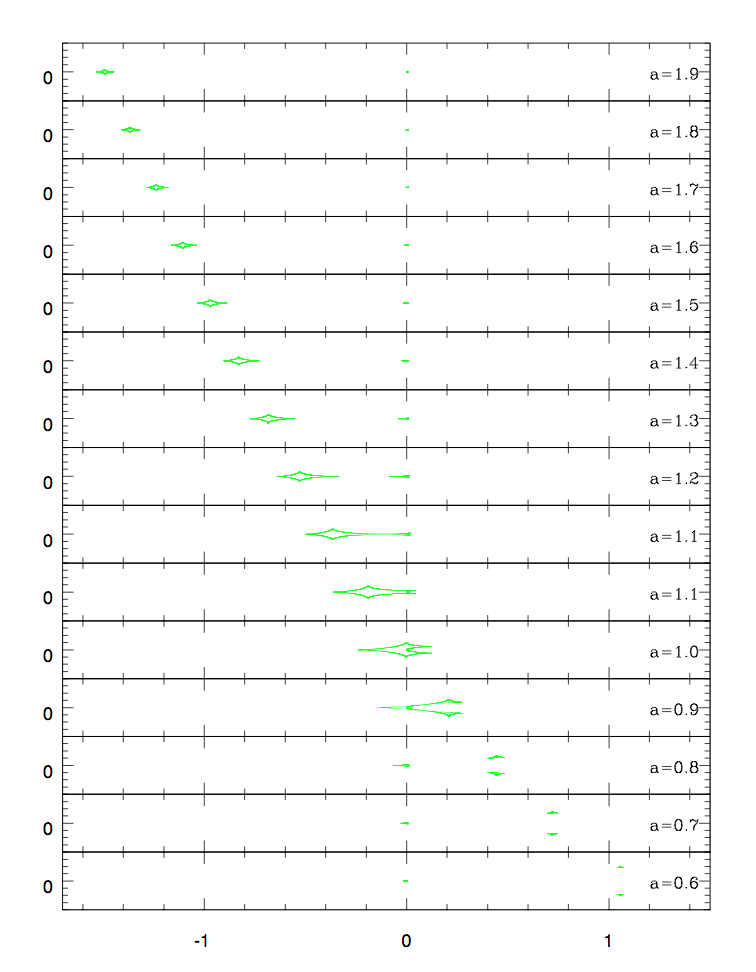}
%created in /data/rd2/rd/gl/gl 
\caption{
Caustic structures for a sequence of values 
%=> for a sequence
of $\alpha: 0.6-1.9$ In each case $q=0.001.$} 
%=> [The caption should also explain what we see on the plot.
%=>  Where is the star? Where is the planet? This is not clear
%=>  to me now...]
%=> [Also, why are there two panels in the figure with a=1.1?]
\end{figure*}

\subsection{Close-Orbit Planets} 

A point lens produces isomagnification contours that are circles centered
at the position of the lens. The addition of a second mass 
changes the isomagnification contours in a way that can be highly
non-linear. In addition, caustic curves are introduced. When the
track of the source passes behind a caustic curve, the number of
images switches from three (five) to five (three); if the source is a point, the
magnification becomes infinite at the time of crossing.
The positions and sizes of the caustics are illustrated in Figure 2
(see also Wambsganss 1997);
the planet used to compute the structures shown in Figure 1 had a mass equal to $0.001$ that
of the central star. The orbital separation in the bottom panel
is $\alpha=0.6.$ 
In this bottom panel we see that the caustic structures are small, so that it
is unlikely that any given source track will encounter one or more of them. 
Furthermore, two of the structures are located more than an Einstein radius
from the central star, in fact much farther from the star than 
is the planet itself. 

Here we will consider close-orbit planets that represent
the small-$\alpha$ continuation of Figure 2. We will call a planet a 
close-orbit planet if $\alpha < 0.5.$ This zone is shown in the left-bottom
panel of Figure 1. Interestingly enough, the detection regime for close-orbit
planets is much larger than the limited zone they occupy. This is 
illustrated in the right-bottom panel of Figure 1, and is explained below.
 
As the value of $\alpha$ decreases, the trend illustrated in
Figure 2 continues. The caustics move out to larger distances from the
star.   
They can be found in a small annulus centered on
 $R_\alpha= \frac{1}{\alpha}-\alpha.$
The caustics themselves are very small, and it is unlikely that the
source track will intersect them. There is, however, a more extended region around
the caustics where the magnification is distorted from the point-lens form.
The size of this region is comparable to but somewhat larger than the
distance between the caustics, which increases as $\alpha$ decreases. 
Isomagnification contours from closer in are pulled out, 
and others from further 
out are pulled in. This means that, for $\alpha\sim (0.2-0.3)$, for example,
 there can be deviations of a few percent that 
start when the source is somewhere between $4.8\, R_E$ and $2.7\, R_E$. These  
are detectable for hours to days,
and can repeat as the perturbed region is whipped around to follow
the motion of the close-orbit planet.  

Thus, the outer portion of the 
close-orbit detection regime starts at slightly larger values of $u$ than $R_\alpha$. 
It generally extends inward 
to the origin, since the reflexive motion of the lens star will repeat,
making  it possible to detect the periodicity and to infer the presence of 
the planet. The discussion above demonstrates that, for close-orbit planets,
the detection regime is significantly larger than the zone within which the
planets lie. details can be found in 
  \rd\~(2012). 
Note again that 
the distinction between the {\sl zone} and the {\sl regime} is that the former
contains the planets, while the latter is the region through which the 
source must pass if we are to detect evidence of the planets, as is 
illustrated in Figure 1.  

\subsection{Resonant-Orbit Planets}

The panels shown in Figure 2 cover the ranges of separations
from just over $0.5\, R_E$ to just under $2.0\, R_E.$
Planets located in this annulus are said to be in the
{\sl zone for resonant lensing} (Mao \& Pacy\'nski 1991; Gould \& Loeb 1992). 
When the planets lie in the zone for resonant lensing, evidence for planets
comes from light curves in which the source tracks pass near or behind the
caustics, producing a significant deviation from the point-lens form.
It is worth noting that the terms ``resonant-orbit'' or 
``zone for resonant lensing'' do not refer to resonance in the
dynamical sense. Instead, the term derives from the circumstance
that discoveries of planets in this zone tend to be associated with
caustic crossings which can produce  
large magnifications that rise and fall over very small time
intervals.

The regime through which the source must pass in order
to detect evidence of planets residing in the zone for resonant
lensing extends from the origin to about $2\, R_E.$ Thus, like the 
close-orbit regime, the resonant regime is larger than the 
zone within which the planets reside.    
The additional area is simply the relatively small area of the disk
that extends from the origin to $0.5\, R_E$. Although small, this region
contains a caustic structure that is associated with the central star.
Thus, the track of any source that passes very close to the star crosses
a caustic and can therefore exhibit evidence of the planet's
presence (Griest \& Safizadeh 1998).   

\subsection{Wide-Orbit Planets}  
If the planet lies in at larger values of $\alpha,$ 
($\alpha$ greater than about $2$),
we will refer to it as a {\sl wide-orbit planet}. 
Wide-orbit planets can serve as almost independent
lenses, and they can also be detected through deviations in
the light curve associated with lensing by the central star
(Di\thinspace Stefano \& Scalzo 1999a, 1999b).

Wide orbit planets can be detected  even for distant approaches, i.e., for 
large values of $\beta.$ The reason is simply that dramatic events can be
created when the source path comes close to the Einstein ring of the planet,
which is, by definition, far from the central star. Such events can occur at
times much earlier than or later than the time of closest approach between the 
foreground and background star. Thus, the wide-orbit detection
zone has an outer edge that is as wide as the widest orbit in the planetary
system serving as lens. 
In addition, evidence of wide-orbit
planets can be detected during close approaches between the
foreground and background star, since the presence of a wide planet can affect
the shape of the stellar-lens light curve. Thus, the wide-orbit
detection regime extends inward to the origin. 
For any given wide-orbit planet, the detection regime may be viewed as a
large-radius annulus, centered on the planet's position, combined with
a disk, centered on the lens star. 
In fact, however, because
a given planetary system may host a sequence of potentially-detectable
wide-orbit planets, it makes sense to view the detection regime as
an extended region, as shown in the upper right panel of Figure 1.

\begin{table}[h]
\centering
\label{tab}
\begin{tabular}{| c | c || c | c |}
\hline
 Regime & $\beta$ &  $\alpha$ & Light Curve \\ \hline
Resonant & $\lesssim2.0$ & $0.5-2.0$ & Figure 7 \\ 
Close & $\lesssim(5.0-6.0)$ & $\lesssim0.5$ & Figures 5 and 6  \\ 
Wide & $all$ & $\gtrsim2.0$ & Figure 4 \\ \hline
\end{tabular}
\caption{
A summary of the properties of the three different zones (defined by $\alpha$)
and regimes (defined by $\beta$).  
$\alpha$ is the orbital separation in units of $\theta_E$ and
$\beta$ is the distance of closest approach in units of $\theta_E$.
See Figure~1. The size of the outer boundary of the close-orbit
regime is determined by the magnitude of the smallest deviations which are
detectable, and can be larger or smaller than $(5-6)\, R_E.$}
\end{table}

\section{Following a Source Star Through the Three Detection Regimes} 

When the foreground star is nearby, its proper motion is the primary source
of relative motion between it and the background star. For the purposes
of this discussion, however, we will work in a frame in which the
foreground star is at rest, and consider the motion of the
background source. We also note that, if the foreground star
is  close enough to us, geometric parallax introduces a measurable
curvature into the relative paths. 
For the purposes of this discussion
we will ignore this effect. 
When, however, we consider a specific
nearby lens, such as VB~10, which is only $5.8$~pc away, parallax
must be explicitly included.

As the source moves toward the foreground star, it first enters the
regime for the detection of wide-orbit planets. If the position of
one or more wide planet happens to be favorable for detection, because
the source will pass close to it, 
planet-lens events can occur long before the source passes close to the
foreground star. In fact, even if the approach between the two
stars is never close enough that the foreground star lenses the
background star in a detectable way, it may be possible to
detect wide-orbit planets. Thus, the wide-orbit regime is explored for many
planetary systems, even if $\beta >>1.$ 

As the source moves closer and closer to the foreground star, planets
in orbits that are ``wide'' but with shorter orbital periods
can also serve as lenses. The orbital motion associated with the shorter
orbital periods increases the probability of an event with a
wide-orbit planet.  

If $\beta$ is smaller than about $5$ or $6,$
the source star enters into the regime for the detection of close-orbit
planets\footnote{The outer radius of the close-orbit regime roughly coincides
with $R_\alpha = \frac{1}{\alpha}-\alpha$, and therefore extends out to values of
$\alpha$ associated with the innermost stable orbit, which generally has $\alpha<<0.1.$
For very small values of $\alpha$, however, there are detectability issues, since the 
deviation of the planet-induced perturbation from baseline is well under a percent.
Here we have somewhat arbitrarily taken the outer edge of the close-orbit
regime to have $R \sim 5-6$, but note that the most appropriate choice depends on the
size of the light-curve effects that can be reliably detected.}.  
%=>  reliably detected.}. [missing period at end of sentence] 
In this regime, the probability     
of detecting the presence of a close-orbit planet approaches unity.
In addition the probability of detecting wide-orbit planets 
increases. 
Thus, even if the value of $\beta$ never reaches a value as small as
$2,$ it is possible to detect planets in a wide range of
orbits, from the widest ($\alpha > 2$) to the closest ($\alpha < 0.5$).

Finally, if $\beta < 2,$ the source enters the resonant-orbit planet
detection regime. While the orbital periods of resonant-orbit
planets are longer than those of close-orbit planets, they can
nevertheless be short enough to significantly increase the 
probability of planet detection. 
In the region $\beta<2,$ planets in all three zones can be detected. 
%=>  $\beta<2,$

The parameters specifying the three zones and regimes are summarized in Table 1. In the next section
we imagine following the track of the source
from the outer edges of the wide-orbit regime to a location very close to the
foreground star.

\subsection{Wide-Orbit Planets}  

\begin{figure*}
\label{iso}
\centering
\includegraphics[width=0.76\textwidth]
{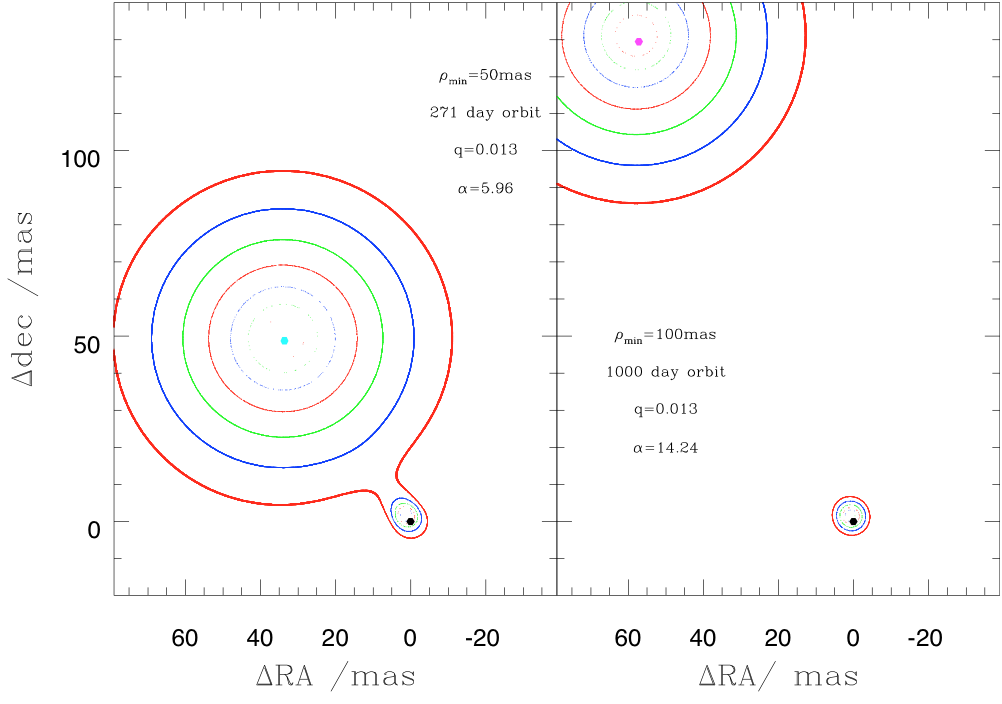}
\caption{
Isomagnification contours for two different wide orbits,
where $\Delta RA = RA- RA_{\rm source}$ and 
$\Delta dec = dec- dec_{\rm source}$. 
 {\sl Left:}~$\alpha=5.96$,
corresponding to a $271$~day orbit, and 
 {\sl Right:}~$\alpha=14.2,$
corresponding to a $1000$~day orbit. We show each system at a time when
the position on the sky of both the planet and the source star is $(0,0).$
The planet produces an independent event
by lensing the source star.  
The isomagnification contours associated with the planet  
can be clearly identified, and the perturbation in both the position
and size of the contours can be seen. The outer red 
contour corresponds to a magnification $(A-1)=0.01$, and the 
magnification associated with each 
contour increases by a one-magnitude increment (a factor of $\sim 2.5$) 
%=> by a one magnitude
as they get closer to the lens. (Note that
the isomagnification contours associated with the planet are
significantly influenced by the central star for $\alpha=5.96$; for the 
wider orbit the effect of central star  on the
region around the planet is smaller.
} 
\end{figure*}

\begin{figure*}
\label{lcwide}
\centering
\includegraphics[width=0.67\textwidth]
{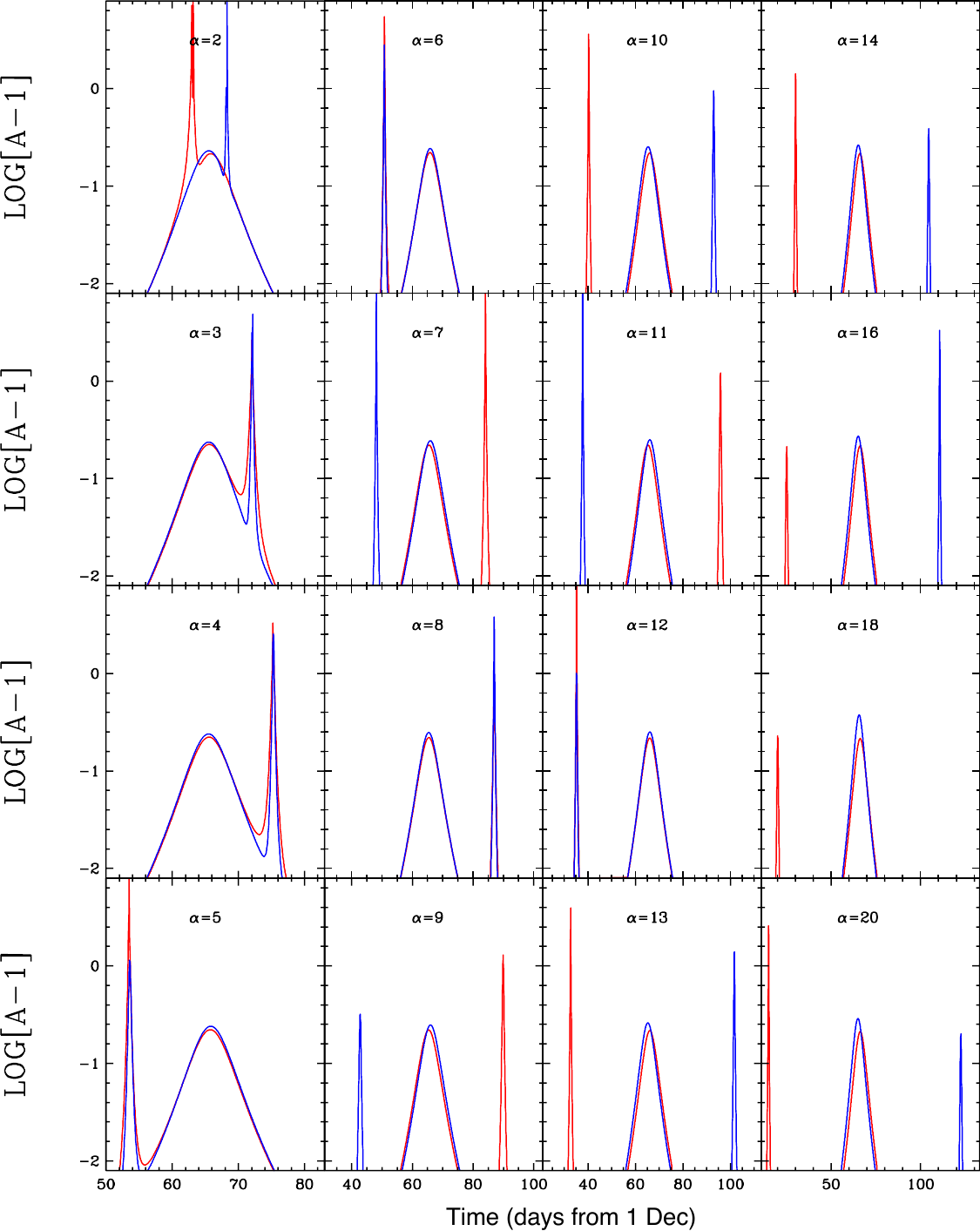}
\caption{
Wide orbit planets, $\beta=1$. Simulated lensing light curves 
%=> Simulated microlensing light curves 
for a variety of orbital separations 
 for a stellar lens orbited by a Jupiter-mass planet. 
Red corresponds to clockwise orbits 
and blue corresponds to counterclockwise.
The wider the orbit, the larger the time between the short-duration
planet-lens event and the more perturbed the stellar-lens event.
This supports the hypothesis that planetary
events can be seen over a wide range of dates depending on the orbital
separation, up to several months before and after the primary
(stellar) event. The date is an arbitrary time prior to the time
of closest passage.} 
\end{figure*}

\subsubsection{Magnification Patterns for Wide-Orbit Planets} 

A star and planet can each produce well-separated events for
projected separations as small as $1.5\, \theta_E,$ 
although this result depends on the value of the mass ratio and on the
photometric sensitivity of the observations. 
In this paper, we consider
the inner boundary of the wide-planet zone to be approximately $2\, \theta_E.$

The wider the planetary orbit, the less likely it is that the track of the
source will pass behind both the planet and the central star (see \rd\ \&
Scalzo 1999b for probability estimates). Nevertheless,  
even when  only the planet-lens event occurs, 
the shape of the light curve
associated with lensing by the planet can be influenced by the 
presence of the star until the value of $\alpha$ is fairly large. 
Figure 3, for example, shows that the $1\%$ isomagnification contour is
significantly distorted from a circular form, even for $\alpha\sim 6.$
The significance of these distortions 
 is twofold: they influence the
light curve characteristics and, because they increase the linear dimensions
of the region within which the magnification is significant, they
increase the event probability. 

Similarly, the isomagnification contours associated with the star can 
be perturbed by the presence of a wide-orbit planet. 
The stellar-lens event may therefore show evidence of the 
planet, even in cases in which the planet does not produce
an independent event. 
This is why the wide-orbit detection regime extends inward to the origin.
In other words, the entire lens plane, 
from the origin out to the largest expected planetary
orbit, comprises the wide-orbit detection regime. 

\subsubsection{Independent Events by Wide-Orbit Planets}

Consider a wide-orbit planet with projected orbital separation $\alpha>2.$ 
Let the distance of closest approach, which will take place at time $t_0$,
be equal to $\beta.$ The path of the source will intersect a circular orbit of radius
$\alpha\, \theta_E$ at two times, given by the following equation. 
\begin{equation} 
%\alpha^2 = 0.160 \Big(t-t_0\Big)^2 + \beta^2 
\Bigg|t-t_0\Bigg| = 
100\, {\rm days}\,  \Bigg(\frac{0.01\, \frac{\theta_E}{{\rm day}}}{\mu}\Bigg)\, 
\Bigg( \alpha^2- \beta^2 \Bigg)^{\frac{1}{2}} 
\end{equation}

Thus, at approximately these two times
it is possible for a wide-orbit planet in a circular orbit of radius $\alpha$ 
to produce a lensing event. The event would last for a time roughly proportional
the square root of the planet mass.\footnote{The planetary Einstein 
radius is smaller than that of the
star by a factor of $\sqrt{M_P/M_\ast},$ 
where $M_p$ is the mass of the planet.} For $\alpha >>2$ the light curve
 would have characteristics
like those expected from an isolated lens. The closer the value of $\alpha$ to the inner edge of the
wide-orbit zone, the more perturbed would be the light curve from the standard point-lens
form.
 
Note that the independent planet-lens event can happen many days
before or after the time of closes passage between the foreground and
background stars.  
Equation 7 highlights an important feature of predicted events, or indeed
any event in which the angle of closest approach is known. {\sl During any
given day before or after the event, we can compute the specific value of
$\alpha,$ the projected orbital separation, of the planet whose lensing
signature we can detect on that day}. 
This is so whenever we know the HPMS's proper motion $\mu,$ and can estimate the
values of $\beta$ and $t_0$. 
Actually, given $\mu, \beta$, and $t_0,$ we can compute a small range
of possible values of $\alpha,$ since the size of the lensing region around
the planet has a finite extent which can be computed for each set of
values of $q$ and $\alpha$ (Figure 3). 
Uncertainties in the path also 
produce uncertainties in the possible day-by-day values of $\alpha$. 
Thus, if an event is detected on a specific day, say the 30th 
day after the main event, we can immediately estimate the value 
of $\alpha$. 
The features of the associated light curve, including the value of $\tau_E,$ 
 allow us to estimate the mass of the planet because the proper motion 
can be well-measured.   

Conversely, if no detection is made on a given day, we know what type of
orbit is {\sl not} producing an event. Lack of evidence for an event
 on any given day 
does not translate to evidence for lack of a planet. It does, however, allow us
to place limits on the possible presence of a planet as follows.
We consider the times the observations that have been made and, 
for each, the minimum value
of the magnification to which   
the observation would have been sensitive. Then, by generating a large number
of orbits for each value of $\alpha$ and each value of $q$, including
a range of orbital eccentricities and inclinations, we can compute what
fraction of the time a planet of each type would be detected.
If the observations would have detected the planet $100\%$ of the time,
then the failure to observe the planet means that it is not there.
If, however, the probability of detecting the planet is ${\cal P},$ then
we can say that there is now a probability of only 
${1-{\cal P}}$ that such a planet 
orbits the lens star.  
This numerical approach can be very accurate, but general planning 
can be facilitated even by less accurate analytic estimates. 
 
A simple analytic approach to computing the 
probability that a wide-orbit planet will
produce a separate short-duration event is to compare
the size of the region within which a planet produces significant effects
with the size of the orbit. This ignores velocity effects, but can nevertheless
provide a reasonable estimate of the fraction, ${\cal P},$ of initial phases 
likely to produce an isolated planet-lens event. 

To compute this fraction, we simply need to compute the ratio between (1)~the
size of the region around the planet that can 
produce detectable lensing effects, 
and, (2)~the size of the orbit. We will express each in units of the
Einstein radius. The denominator, the size of the orbit, is simply 
$2\, \pi\, \alpha$.  Developing a good approximation
for the value of the numerator requires some care. First, the size of the
region within which lensing can be detected depends on the
sensitivity of the observations. If only magnifications of $34\%$
($6\%, 1\%$)  or more are
detectable, then the width of the lensing region is 
$1 \times (2  \,  R_E)$ [$2  \times (2\, R_E), 3.5 \times  (2\, R_E$)]. 
The factor to the left in each case corresponds to the distance, in Einstein
radii, at which the magnification becomes detectable. 
We have taken this into account  
by introducing a factor $F/2$, which is unity when  
deviations of $6\%$ are detectable, $0.5$ if deviations must be
larger than $34\%$ are detectable, $3.5/2=1.75$ if deviations must be 
larger than only $1\%$ to be detectable.  

A second factor influences the size of the lensing region associated with the
planet. This is the size of the orbit. For values of $\alpha$ in the range
$2-6$, the lensing region is significantly stretched, with the amount 
of stretching also dependent on the mass ratio, $q$. 
  For larger values of
$\alpha$ the stretching is minimal. To take this into account, we
have introduced the factor $E(q_i,\alpha)/1.5$, which has the value 
roughly equal to $1$ for $\alpha=3$ and $q=0.001.$    

%\newpage
With these definitions, 
the probability, ${\cal P}$, that a wide-orbit planet will produce an isolated event is  
\begin{equation} 
{\cal P}=0.36\, \frac{F}{2} \sum\limits_{i=1}^n \Big(\frac{q_i}{0.01}\Big)^{\frac{1}2{}}
\, \Big(\frac{1}{\alpha_i}\Big)\, \Big(\frac{E(q_i,\alpha_i)}{1.5}\Big),    
\end{equation} 
where the sum is over wide-orbit planets. For $\alpha=3,$ the
contribution is $0.12.$ 
A more accurate estimate of the probability
would include the effects of orbital motion and parallax.

\subsubsection{Detecting evidence of the planet during the stellar-lens event}
Fits to light
curves produced by the star have the potential to
identify deviations from the point-lens form
and to discover 
 wide-orbit planets. {\sl This is the case even when the 
planet itself does not produce a separate detectable event.}

The influence of the planet could be manifest through a slight
distortion of the isomagnification contours surrounding the star.
Such effects are likely to be most prominent in the low-magnification 
portion of the light curve.

The planet could also be detected through the reflexive motion of the star.
As shown in Equation 4,
the magnitude of the positional shift is small for small values of $\alpha.$
Fortunately, for small $\alpha$, direct detection of the planet is 
more likely (Equation 8). Also for small $\alpha$, if the relative speed is not too high,
the phase could change significantly enough during a close approach to
affect the light curve shape. The wider the orbit, the larger the 
reflexive motion, and the more likely it is to produce deviations in the
stellar-lens light curve from the pure point-lens form. This is illustrated 
in Figure 4.   

\newpage
\subsection{Close-Orbit Planets}

\begin{figure*}
\label{pathswide}
\centering
\includegraphics[width=0.7\textwidth]
{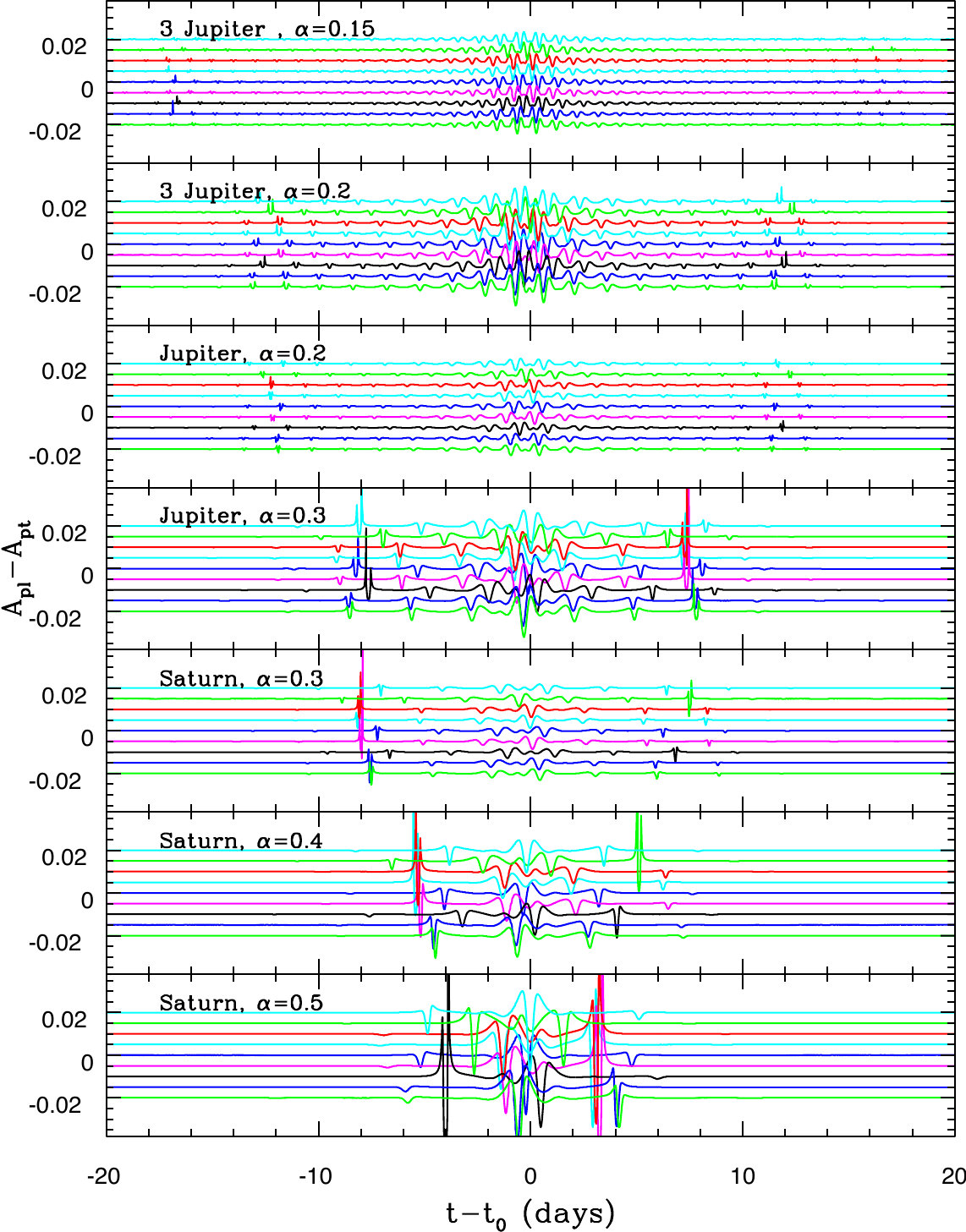}
\caption{
The difference between planet-lens ($A_{pl}$) and point-lens ($A_{pt}$)
behavior as a function of $t-t_0$ for close-orbit planets.
This illustrates planet effects
for a range of orbital separations ($\alpha$) and masses on the $5$~mas
approach ($\beta=0.5$).
Light curves from different initial orbital phases are shown with
vertical offsets for clarity.
Small, quasiperiodic deviations from single lens behaviour
are observed, and there is increased activity in the region
of $u\sim1/\alpha-\alpha$, near the beginning and end of the
variability pattern.
The deviations close to $t_0$ are caused by reflexive motion of the central star.
}
\end{figure*}

\begin{figure*}
\label{pathswide}
\centering
\includegraphics[width=0.7\textwidth]
{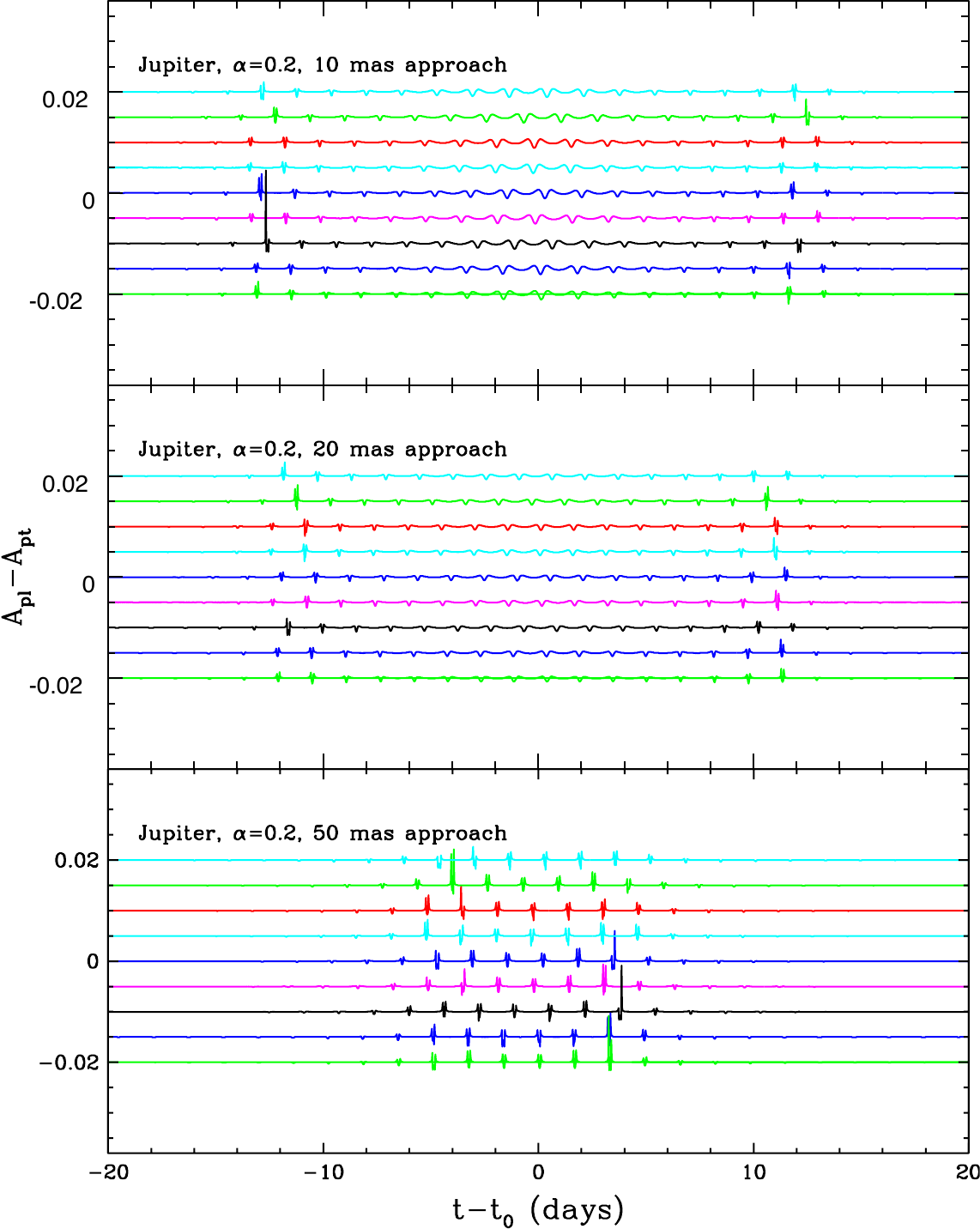}
\caption{
The difference between planet-lens ($A_{pl}$) and point-lens ($A_{pt}$)
behavior as a function of $t-t_0$, analogously to Figure 4. This illustrates close-orbit planet effects
for $\alpha=0.2$, for a Jupiter-mass planet and
$b=5,10,20$~mas ($\beta=1,2,5$).
 The light curves from different phases are plotted with vertical
 offsets for clarity
so the y axis should be seen as a relative scale only.
Small, quasiperiodic deviations from single lens behaviour
are produce, and there is increased activity in the region
of $u\sim1/\alpha-\alpha$, near the beginning and end
of the periodic sequence.
The deviations close to $t_0$ are caused by reflexive motion of the central star.
}
\end{figure*}

If the foreground star happens to have planets with orbits much 
smaller than the Einstein
radius, the lensing signatures can be significantly perturbed from the
point-lens form.
This is because there is a small region within which the
isomagnification contours are distorted from the circular shape they would have 
had in the absence of the planet. These perturbations are 
maximized at a distance 
$R_\alpha = (\frac{1}{\alpha}-\alpha)>1.5$ from the center of mass.
In the absence of orbital motion, 
there would be only a small
probability that the distorted region
 would pass in front of the source. 
The orbital period is short for small values of $\alpha,$ however, especially
when the planetary system serving as a lens is nearby.
Thus, even though the deviations tend to be small, repetition confers an
advantage. 

The light curves associated with close-orbit planets have the same overall shape
as light curves produced by the central star. To detect evidence of the
planets, we must be sensitive to small deviations of the type studied
in Di\thinspace Stefano (2012). The deviations have a characteristic 
up-down-up-down
shape, and for nearby lenses, will generally repeat. Figures 3 through 6 of
Di\thinspace Stefano 2012 illustrate the deviations in the isomagnification
contours and light curves for a ``hot Jupiter'', and for a Neptune-mass and an
Earth-mass planet located in the habitable zone of a dwarf star. In Figures 5
and 6 of this paper, we show the light curves for a variety of planets and planetary orbits,
plotting the deviations, $A_{pl}-A_{pt}$, 
from the point-lens light curve produced by the central 
star alone. The specific foreground star for which
the calculations were done is VB~10; the central star is, therefore 
a dwarf star of mass $0.075\, M_\odot$. The value of $R_E$ is roughly 
$0.08$~AU. Here we explain the patterns.

The most distinctive features are repeating deviations that start in the 
wings (i.e.,
the low-magnification portion) of the light curve. For each
value of $\alpha,$ the deviations start as the source star approaches
a distance $R_\alpha$ from the central star. The deviations
are larger for larger values of $\alpha.$ In fact, if the observations
are able to reliably detect deviations of size $\delta,$ then close
orbit planets with semimajor axes in the range between $\alpha_{small}$
and  $0.5$ 
can be detected, where
\begin{equation}  
\alpha_{small}=0.84\, \delta^{\frac{1}{4}}.  
\end{equation}
This effect is illustrated in the early-time and late-time
deviations seen in the light curves of Figures 5 and 6, where the
size and location of the planet-induced perturbations clearly 
depend on the value of $\alpha.$

Detectability is determined not just by 
size of the perturbations, 
but also by their duration. The  
duration depends on the value of the orbital period and 
on the mass ratio.
While shorter orbital periods increase detectability by
increasing the number of repetitions, the faster motion associated with them
decreases the duration of perturbations. On the other hand, increasing the
mass ratio $q$, between the planet and star always increases the
duration. Let $T_{dev}$ represent the duration of perturbations. \rd\ (2012) shows that 
\begin{equation}  
T_{dev}=
%\frac{L(\alpha, q)}{2\, \pi\, R_\alpha}\, P_{orb} = 
2.5\, \xi\, P_{orb} 10^{[0.5\, Log_{10}(q) - 0.2]}, 
\end{equation}
where the value of $\xi$ depends on the observing setup, and can be
taken to be of order unity in this discussion. (See \rd\ 2012 for more details.)   
Figure~5 displays the early-time and late-time perturbations associated
with close-orbit planets. 
It shows that 
like wide-orbit planets,  
close-orbit planets can produce deviations both 
in advance of 
and after
the closest approach. The time at which the deviations start is
the time at which the source star enters the region with magnification
perturbations at $R_\alpha.$ 
Let $t_{max}$ be the time at which deviations related to 
the planet begin (before the time of closest passage) or end (after the time of closest passage),
representing the maximum time difference between the time of closest approach 
and the time at which deviations occur.  If we measure time in days we 
have the analogue to Equation (3) for close-orbit planets:
\begin{equation} 
\Bigg|t_{max}-t_0\Bigg| \sim 
100\,{\rm days}\,  \Bigg(\frac{0.01\, \frac{\theta_E}{{\rm day}}}{\mu}\Bigg)\, 
\Bigg( \Big(\frac{1}{\alpha} - \alpha \Big)^2 - \beta^2 \Bigg)^{\frac{1}{2}} 
\end{equation}

The smaller the value of $\alpha,$ the larger the value of $R_{\alpha},$ 
and the longer the difference
in time between the closest approach and the start of close-planet-lens 
deviations. This is
shown in Figure~5, which plots the difference between the magnification
which includes the effect of the planet, $A_{pl},$ and 
the point-lens magnification, $A_{pt}$, expected if there is no planet.  
Figure~5 illustrates that the effects diminish
in size as the value of $\alpha$ becomes smaller. Thus the effects of a 
3-Jupiter-mass planet at $\alpha=0.15$ would be challenging for ground-based
observations, while a Saturn-mass planet could easily be discerned for
$\alpha$ larger than about $0.3.$ The early-time and late-time perturbations
are also shown in the top two panels of Figure~6.

The bottom panel of Figure~6 illustrates that the deviations produced
by close-orbit planets can be detected even when the distance of closest
approach is larger. The example shown in this panel has $\beta=5.$    
It is interesting to note that $A_{pt}$ reaches a maximum
value of only $1.00275$. Thus, if we had plotted only the raw
magnification, $A_{pl}$, the light curve would be essentially the
same as the one shown. Thus, the signature of the planet would 
be an oscillatory deviation from baseline, of finite duration. 

Another similarity with the wide-orbit case is that
we can compute a specific value of $\alpha$ for a planet whose 
lensing signature we are capable of detecting (given the sensitivity
and cadence of sampling) on a given day. Thus, when we do not find evidence
of planets, we can place quantifiable limits on their existence.

Note that, by selecting VB~10 as an example, we have chosen a star with
particularly high proper motion. Such stars are the most likely to
produce events that can be predicted. The high proper motion does decrease the
time duration of deviations, which makes detection more challenging. Other cases
are shown in \rd\ (2012) and in a future paper, based on \rd , Matthews, \& L\'epine (2012). 

In addition to the repeating small deviations in the wings of the 
light curves, Figure~5 and the top two panels of Figure~6 
show another interesting effect. This is
that,  for small values of $\beta,$ there is a second region displaying
significant perturbations. This region is near the light curve's peak, when the background
star makes its closest approach to the foreground star.
It corresponds to the reflexive motion of the central star and is larger for
larger orbits.  
The size of the reflexive effects
also increases with the mass of the close-orbit planet.  

Just as there is a largest orbital separation for a wide-orbit planet, there is
a smallest orbital separation for a close-orbit planet.  
To avoid catastrophic tidal disruption a planet must satisfy
 the condition $r_h\gtrsim2r_p$, where $r_h$ is
 the Hill radius given by $r_h\approx a(\frac{m_p}{3M_*})^{1/3}$. Thus
 the minimum orbital separation of a planet around VB~10 is 
roughly given by 
\begin{equation} 
a_{min}\sim 0.3R_\odot \frac{r_p}{r_J}\left(\frac{M_*}{0.075M_{\odot}}\right)^{1/3}\left(\frac{0.001M_{\odot}}{m_p}\right)^{1/3}
\end{equation}
This corresponds to $\alpha_{min} \sim 0.03$ for the VB~10 case.
The deviations produced by a planet in such a close orbit  would occur when 
the position of the source star on the sky was 
approximately $30\, R_E$ from the position of the foreground star. 
Their magnitude would be so small that they would be difficult
to detect with today's technology.

\noindent{\bf Event Probabilities:} In Figures 5 and 6, each panel 
contains 9 independent light curves, slightly
displaced from each other so that the variations in each can
be resolved. The nine light curves in each panel differ from 
each other only in that 
a different initial phase was chosen for each. {The effects were similar
in all cases.} Had the perturbations been potentially detectable for only
a small fraction of initial phases, then the probability of
planet detection would be comparable to the value of that fraction.
The ubiquity of detectable results 
indicates that, if the observational strategy
is capable of detecting deviations caused by planets with a given value of 
$\alpha$, orbital period, and $q$, then it will detect such 
planets with certainty. Thus, either close-orbit
planets will be discovered, or limits on their existence can always
be placed, if the observing conditions are favorable and the monitoring program
is well designed. The values of $\alpha$ for which we can derive limits or else
guarantee a detection are determined by the photometric sensitivity, whilst 
those of $q$ are 
determined by the cadence of sampling.

\subsection{``Resonant'' Regime} 

\begin{figure*}[!ht]
\label{lc_caust}
\centering
\includegraphics[width=0.7\textwidth]
{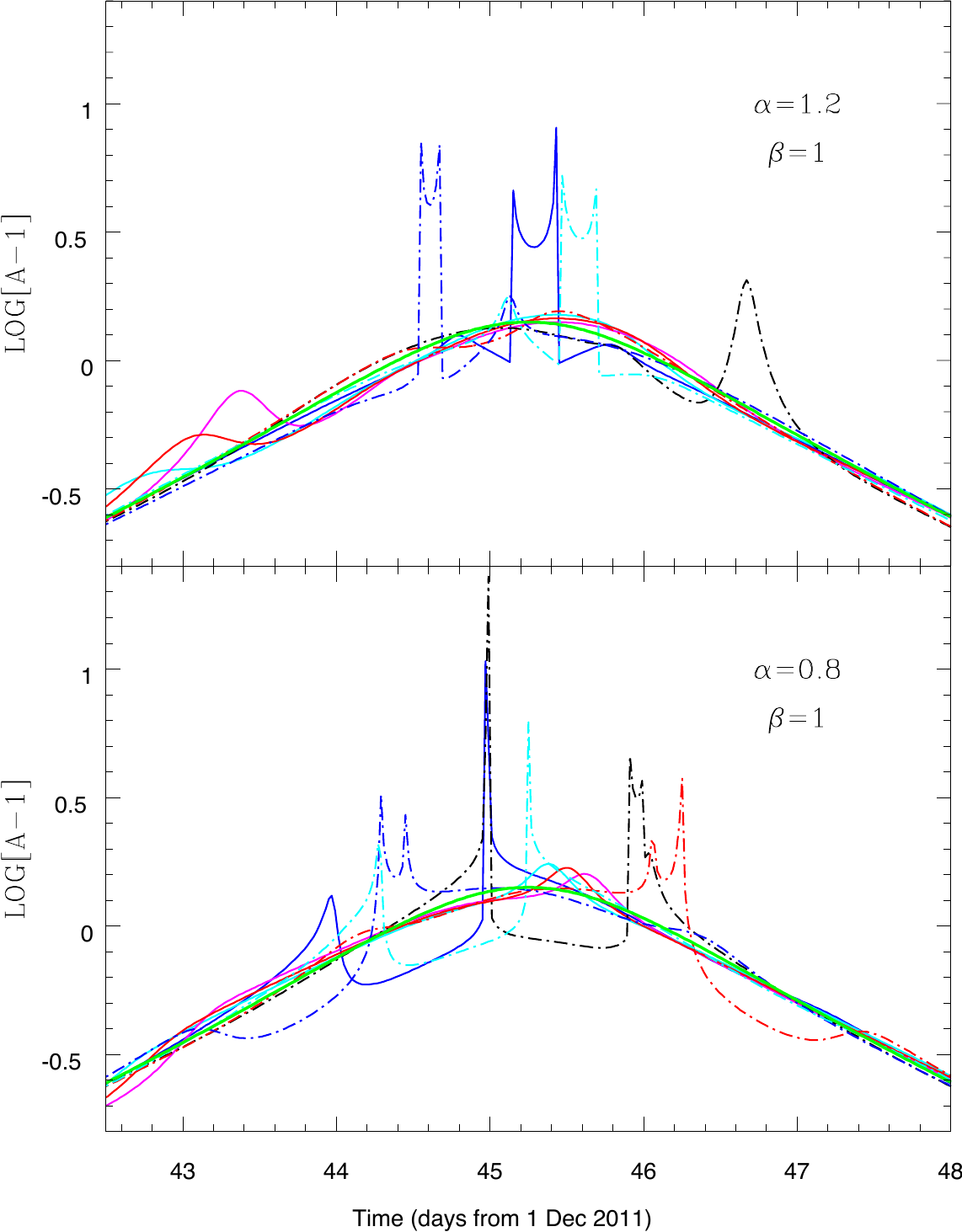}
\caption{
Simultaneously plotted light curves for a range of randomly sampled initial phases 
(in different colours), $\phi_0$,
for two different values of $\alpha$ on a path that approaches within one Einstein radius of the 
central star. These light curves illustrate
the types of features associated with lensing in the resonant zone.
Caustic crossings can cause very sharp
increases in magnification that
are easily detectable, but the smoother curves demonstrate that significant deviations
can occur even when caustics are not crossed.
Ten separate light curves are plotted in each panel. The fact that a
large fraction of them exhibit high magnifications and/or significant deviations
from the point-lens form shows that the probability of detectable
planet-lens distortions is large.
}
\end{figure*}

When the possibility for discovering planets via microlensing was first
discussed (Mao \& Paczy\'nski 1991; Gould \& Loeb 1992),
attention focused on 
passages of the source star near or behind caustic structures like
those shown in Figure 2. The associated light curves can show dramatic effects,
although these can be moderated by the smoothing effect associated with
the finite size of the source.    
Until now, microlensing 
searches for planets have focused on finding planets in the resonant zone.
Lensing events are found by observing teams monitoring wide fields. 
Events judged
to have a high probability of exhibiting planet-lens signatures are targeted for 
almost continuous monitoring during an interval (which can be as
short as hours) when the 
magnification is deemed likely to be significantly influenced by the
presence of planets.
 The $16$ planets discovered via
lensing have been found through such methods and are squarely in or near the
borders of the zone for resonant lensing.  

\noindent {\bf Light Curves:} 
The associated light curves are therefore familiar from both calculations 
and observations. The new element introduced when the lens is a nearby foreground
star is that the orbital size for values of $\alpha$ in the resonant zone 
can be small enough to make the orbital periods relatively short. Thus, rotation
can play a significant role in determining the form of the light curves.
This is illustrated in Figure 7. Because of the effect of orbital motion, the individual light curves plotted there
exhibit more structure than typical for more distant lenses with planets
in the zone for resonant lensing.    

\noindent{\bf Event Probabilities:}
The closer the approach, the higher the probability of detecting events in which
the source has passed near a caustic structure. This is 
illustrated in Figure 7, where light curves for a $5$~mas approach
%=> illustrated in Figure 3,
of VB~10 to a background star are shown.
For $\alpha=1.2$ (top panel) and
$\alpha=0.8$ (bottom panel), ten randomly selected initial phases were each used to
generate a light curve. For $\alpha=1.2,$ eight of the light curves showed
significant deviations, three involving caustic crossings (those exhibiting wall-like
rises and falls). For $\alpha=0.8,$ all ten light curves displayed significant variations,
with an even larger number of caustic crossings than for $\alpha=1.2$.
This illustrates that, for close passages between a background star and a nearby 
planetary system, the probability of detecting any planets inhabiting the zone for resonant
lensing is close to unity.   

\newpage
\section{Conclusions}

We have studied, for the first time, the characteristics and timing
of planet-influenced events that can occur when a nearby foreground 
star makes a predicted close approach to a background star. Perhaps the
%=> to a background
most interesting finding is that the probability
of discovering planets can be high enough to make monitoring
worthwhile even when the passage between the foreground and
background star is not that close.
This is because the Einstein rings of nearby lenses can be large 
(from around a few~mas to $30-40$~mas), while close orbit planets,
should they exist, 
are almost certain to produce detectable events  
for approaches of $(5-6)\, \theta_E$ or closer. These events could occur 
weeks or months before (or after) the time of closest approach. 
In addition, wide-orbit planets can produce events 
%=> In addition
for even more distant approaches, and at even earlier or later times,
although with smaller probability. The detection of planets
in the resonant zone
requires the closest approaches and is therefore most likely 
just near the time of closest approach, during an interval when the
magnification reaches its peak value.     

In order to relate the type of event to the time at
which it is expected to occur (relative to the time of closest approach),
we have introduced the notion of {\sl detection regimes}, the regions within
which it is most likely to detect planets in each zone. 
The close-orbit
detection regime is the region through which the source must 
pass in order for us to detect close-orbit planets ($\alpha < 0.5$).
The regime contains 
the zone where the
planets are located, but also extends well beyond it.
The {\sl resonant regime} contains a disk encompassing the central star,
as well as the annulus that constitutes the
zone for resonant  
lensing ($0.5 < \alpha < 2$). The wide regime, encompasses the entire
lens plane, from the central star out to the radius of the farthest planet.    
Wide orbit planets can be detected at very early and/or at very
late times and for distant approaches; the light curves are similar to
the light curves expected for low-mass isolated lenses. In addition, if the
approach is close enough to generate a stellar-lens event, the reflexive
motion caused by a distant planet can be magnified, influencing 
the light curve shape, even if the
wide-orbit planet does not itself generate an event.  
Close-orbit planets can be detected at intermediate
times and for approaches less than about $6\, \theta_E.$ The light curve
signatures are characteristic up-down-up-down features whose duration 
and baseline
magnification can be used to determine the characteristics of the planet
and its orbit.
Resonant-orbit planets can be detected only   
during close approaches. When the lens is nearby enough for a close passage to be predicted, orbital motion can increase the structure in the light curve and
thus increase the probability of planet detection.

These theoretical results provide a good basis for the design of observing
programs that can effectively take advantage of close passages between
nearby high-proper-motion stars and background stars. In 
a future paper, based on \rd , Matthews, \& L\'epine (2012),
we specifically address the design of optimal programs, with the goal of
opening up a new and productive approach to the discovery and study of
nearby planets.  
\bigskip

\noindent {\bf Acknowledgements}

\smallskip
%\newpage
\noindent We thank Christopher Stubbs; Christopher Crockett; Fred Walters;  
David Charbonneau, Zachory Berta and the MEarth project; 
Jochen Greiner at GROND; Matthew Templeton and the AAVSO for their help 
and advice. This work was supported in part by NSF under 
AST-0908878 and under AST-0908406. 
%=> Please also mention my NSF grant, AST-0908406.

%\end{acknowledgements}

\bigskip
%\newpage


\begin{thebibliography}{}
\bibitem[Aihara et al.(2011)]{2011ApJS..193...29A} Aihara, H., Allende 
Prieto, C., An, D., et al.\ 2011, \apjs, 193, 29 

%\bibitem[Anglada-Escude et al.(2010)]{anglada2010}
%Anglada-Escude, et al. 2010, \apj, 711, L24

\bibitem[Batista2011]{batista2011}
Batista, V., et al. 2011, \aap, 529, 102

%\bibitem[Bean et al.(2010)]{Bean2010}
%Bean, J. L., et al. 2010 \apj, 711, L19

%\bibitem[Bolatto Falco (1994)]{Bolatto.Falco.(1994)}
%Bolatto, A. D., \& Falco, E. E., 1994, \apj, 436, 112

\bibitem[Bond et al.(2001)]{2001MNRAS.327..868B} Bond, I.~A., Abe, F., 
Dodd, R.~J., et al.\ 2001, \mnras, 327, 868 

\bibitem[Di Stefano(1999)]{1999ApJ...512..558D} Di Stefano, R.\ 1999, \apj,
512, 558

\bibitem[DiStefano(2008a)]{DiStefano.2008a}
Di Stefano, R., 2008a, \apj, 684, 46

%Discovering Habitable Earths, Hot Jupiters, and Other 
%Close Planets with Microlensing
\bibitem[DiStefano_close.2012]{DiStefano.2012}
Di Stefano, R., 2012, \apj, 752, 105  

\bibitem[DiStefano(2008b)]{DiStefano.2008b}
Di Stefano, R., 2008b, \apj, 684, 59


\bibitem[rd_etal_2012]{rd_matthews_lepine.2012}
Di Stefano, R., Matthews, J., L\'epine, S. 2012, arXiv:1202.5314

\bibitem[Di Stefano \& Night(2008)]{2008arXiv0801.1510D} 
Di Stefano, R., \& Night, C.\ 2008, arXiv:0801.1510 

\bibitem[Di Stefano \& Scalzo(1999)]{1999ApJ...512..564D} 
Di Stefano, R., \& Scalzo, R.~A.\ 1999a, \apj, 512, 564 

\bibitem[Di Stefano \& Scalzo(1999)]{1999ApJ...512..579D} 
Di Stefano, R., \& Scalzo, R.~A.\ 1999b, \apj, 512, 579  

\bibitem[Dominik 
\& Sahu(2000)]{2000ApJ...534..213D} Dominik, M., \& Sahu, K.~C.\ 
2000, \apj, 534, 213 

\bibitem[Einstein(1936)]{1936Sci....84..506E} Einstein, A.\ 1936, 
Science, 
84, 506 

%\bibitem[Feibelman, W. A. (1966)]{1966Sci...151...73F}
%Feibelman, W. A. 1966, Science 151, 73

%\bibitem[Feibelman, W. A. (1986)]{1986PASP...98.1199F}
%Feibelman, W. A. 1986, \pasp 98, 1199

\bibitem[Guo et al.(2011)]{Guo et al.(2011)}
Guo, X., et al. 2011, arXiv:1112.4608G

\bibitem[Gould \& Loeb(1992)]{Gould.Loeb.1992}
Gould, A., \& Loeb, A. 1992, \apj, 396, 104

%\bibitem[Greiner et al. (2008)]{Greineretal.(2008)}
%Greiner, J., et al. 2008 \pasp, 120, 405

\bibitem[Griest \& Safizadeh(1998)]{Griest.Safizadeh.1998}
Griest, K., \& Safizadeh, N. 1998, \apj, 500, 37

%\bibitem[Hilton et al. (2010)]{Hilton2010}
%Hilton, E. J., et al. 2010,\aj, 140, 1402 

%\bibitem{mearth}
%Irwin, J., et al. 2008, arXiv:0807.1316v1

\bibitem[Kaiser (2002)]{Kaiser.2002}
Kaiser, N. 2004, SPIE, 5489, 11
%\bibitem[Lazorenko et al.(2011)]{Lazorenko.etal.2011} 
%Lazorenko, P. F., et al. 2011, \aap, 527, 25

\bibitem[L{\'e}pine 
\& DiStefano(2012)]{2012ApJ...749L...6L} L{\'e}pine, S., \& DiStefano, R.\ 2012, \apjl, 749, L6 


\bibitem[L{\'e}pine 
\& Gaidos(2011)]{2011AJ....142..138L} L{\'e}pine, S., \& Gaidos, E.\ 2011, \aj, 142, 138 

\bibitem[L{\'e}pine 
\& Shara(2005)]{2005AJ....129.1483L} L{\'e}pine, S., \& Shara, M.~M.\ 2005, \aj, 129, 1483 


\bibitem[Mao \& Paczynski(1991)]{Mao.Paczynski.1991}
Mao, S., \& Paczynski, B. 1991, \apj, 374, 37

%\bibitem[Mayor et al.(2010)]{Mayor.etal.2010}
%Mayor, M., et al. 2010, \aap, 507, 487

%\bibitem[Paczynski(1995)]{Paczynski.1995}
%Paczynski, B. 1995, Acta Astronomica, 45, 345

%\bibitem[Penny et al.(2011)]{2011MNRAS.417.2216P}
%Penny, M. T. et al. 2011, \mnras 417, 2216 

%\bibitem[Pravdo \& Shaklan(2009)]{Pravdo.Shaklan.2009}
%Pravdo, E., \& Shaklan 2009, \apj, 700, 623

%\bibitem[Salim \& Gould(2000)]{Salim.Gould.2000}
%Salim, S., \& Gould, A. 2000, \apj, 539, 241

\bibitem[Udalski(2003)]{2003AcA....53..291U} Udalski, A.\ 2003, \actaa, 53, 
291 


%\bibitem[Van Biesbroeck(1944)]{VanBiesbroek.1944}
%Van Biesbroeck, G. 1944, \aj, 51, 61

\bibitem[Wambsganss(1997)]{1997MNRAS.284..172W} Wambsganss, J.\ 1997, 
\mnras, 284, 172 



%\bibitem[West et al. 2008]{Westetal.2008}
%West, A. A., et al. 2008, ApJ, 135, 785

%\bibitem[Zapatero Osorio et al.(2009)]{2009A&A...505L...5Z}
%Zapatero Osorio, M. R. et al. 2009, \aap, 505, L5

\end{thebibliography}
\end{document}